\documentstyle[psfig,times]{mn}  
\begin{document}
\def\reef{\par\noindent\hang}
\def\etal{et al.\ }
\def\eg{{\em eg.\ }}
\def\etc{{\em etc.\ }}
\def\ie{{\em i.e.\ }}

\def\spose#1{\hbox to 0pt{#1\hss}}
\def\approxlt{\mathrel{\spose{\lower 3pt\hbox{$\sim$}}
	\raise 2.0pt\hbox{$<$}}}
\def\approxgt{\mathrel{\spose{\lower 3pt\hbox{$\sim$}}
	\raise 2.0pt\hbox{$>$}}}
	
\def\Mdot{\hbox{$\dot M$}}
\def\degmark{$^\circ$}
\def\<{\thinspace}
\def\s{\hbox{\phantom{5}}}	%one space
\def\ss{\s\s}		%two spaces
\def\sss{\ss\s}		%three
\def\ssss{\ss\ss}	%four
\def\lit{\obeyspaces\obeylines}
%
% 	Simple units
\def\arc{{\rm\thinspace arcsec}}
\def\cm{{\rm\thinspace cm}}
\def\ct{{\rm\thinspace ct}}
\def\erg{{\rm\thinspace erg}}
\def\eV{{\rm\thinspace eV}}
\def\g{{\rm\thinspace g}}
\def\G{{\rm\thinspace G}}
\def\ga{{\rm\thinspace gauss}}
\def\K{{\rm\thinspace K}}
\def\keV{{\rm\thinspace keV}}
\def\m{{\rm\thinspace m}}
\def\km{{\rm\thinspace km}}
\def\kpc{{\rm\thinspace kpc}}
\def\Lsun{\hbox{$\rm\thinspace L_{\odot}$}}
\def\rad{{\rm\thinspace rad}}
\def\MeV{{\rm\thinspace MeV}}
\def\Mpc{{\rm\thinspace Mpc}}
\def\Msun{\hbox{$\rm\thinspace M_{\odot}$}}
\def\pc{{\rm\thinspace pc}}
\def\ph{{\rm\thinspace photons}}
\def\s{{\rm\thinspace s}}
\def\yr{{\rm\thinspace yr}}
\def\sr{{\rm\thinspace sr}}
\def\Hz{{\rm\thinspace Hz}}
\def\GHz{{\rm\thinspace GHz}}
\def\W{{\rm\thinspace W}}
%	Compound units
\def\cmps{\hbox{$\cm\s^{-1}\,$}}
\def\ctps{\hbox{$\ct\s^{-1}\,$}}
\def\cmsq{\hbox{$\cm^2\,$}}
\def\cmcu{\hbox{$\cm^3\,$}}
\def\pHz{\hbox{$\Hz^{-1}\,$}}
\def\pcmcu{\hbox{$\cm^{-3}\,$}}
\def\ergcmcups{\hbox{$\erg\cm^3\ps\,$}}
\def\ergpcmps{\hbox{$\erg\cm^{-3}\s^{-1}\,$}}
\def\ergpcmsqps{\hbox{$\erg\cm^{-2}\s^{-1}\,$}}
\def\ergpspcmsq{\hbox{$\erg\cm^{-2}\s^{-1}\,$}}
\def\ergpspkpcsq{\hbox{$\erg\s^{-1}\kpc^{-2}\,$}}
\def\ergpspA{\hbox{$\erg\s^{-1}\AA^{-1}\,$}}
\def\ergpspcmsqpA{\hbox{$\erg\s^{-1}\cm^{-2}$\AA$^{-1}\,$}}
\def\ergpcmsqpspsqarcsec{\hbox{$\erg\cm^{-2}\s^{-1}\arc^{-2}\,$}}
\def\ergpcmsqpspapsqarcsec{\hbox{$\erg\cm^{-2}\s^{-1}\AA^{-1},\arc^{-2}\,$}}
\def\ergps{\hbox{$\erg\s^{-1}\,$}}
\def\gpcm{\hbox{$\g\cm^{-3}\,$}}
\def\gpcmps{\hbox{$\g\cm^{-3}\s^{-1}\,$}}
\def\gps{\hbox{$\g\s^{-1}\,$}}
\def\WpHz{\hbox{$\W\Hz^{-1}\,$}}
\def\kmps{\hbox{$\km\s^{-1}\,$}}
\def\ksec{\hbox{$ksec\,$}}
\def\Lsunppc{\hbox{$\Lsun\pc^{-3}\,$}}
\def\Msunpc{\hbox{$\Msun\pc^{-3}\,$}}
\def\Msunpkpc{\hbox{$\Msun\kpc^{-1}\,$}}
\def\Msunppc{\hbox{$\Msun\pc^{-3}\,$}}
\def\Msunppcpyr{\hbox{$\Msun\pc^{-3}\yr^{-1}\,$}}
\def\Msunpyr{\hbox{$\Msun\yr^{-1}\,$}}
\def\pcm{\hbox{$\cm^{-1}\,$}}
\def\pcmsq{\hbox{$\cm^{-2}\,$}}
\def\pcmsqpkeVps{\hbox{$\cm^{-2}\keV^{-1}\s^{-1}\,$}}
\def\pmsq{\hbox{$\m^{-2}\,$}}
\def\radpmsq{\hbox{$\rad\m^{-2}\,$}}
\def\pcmcuK{\hbox{$\cm^{-3}\K$}}
\def\phps{\hbox{$\ph\s^{-1}\,$}}
\def\phpcmsqps{\hbox{$\ph\cm^{-2}\s^{-1}\,$}}
\def\pHz{\hbox{$\Hz^{-1}\,$}}
\def\pMpc{\hbox{$\Mpc^{-1}\,$}}
\def\pMpccu{\hbox{$\Mpc^{-3}\,$}}
\def\ps{\hbox{$\s^{-1}\,$}}
\def\psqcm{\hbox{$\cm^{-2}\,$}}
\def\psr{\hbox{$\sr^{-1}\,$}}
\def\pyr{\hbox{$\yr^{-1}\,$}}
\def\kmpspMpc{\hbox{$\kmps\Mpc^{-1}$}}
\def\Msunpyrpkpc{\hbox{$\Msunpyr\kpc^{-1}$}}

\title{Infrared observations of serendipitous hard {\sl Chandra} X-ray sources} 

\author[C.S. Crawford et al ]
{\parbox[]{6.in} {C.S. Crawford, A.C. Fabian, P. Gandhi, R.J. Wilman
and R.M. Johnstone
\\
\footnotesize
Institute of Astronomy, Madingley Road, Cambridge CB3 0HA \\}}

\maketitle
\begin{abstract}
We present observations of a sample of optically-faint, hard X-ray
sources of the kind likely to be responsible for much of the hard
X-ray background. We confirm that such sources are easily detected in
the near-infrared, and find that they have a featureless continuum
suggesting that the active nucleus is heavily obscured. The infrared
colours of the majority of the targets observed are consistent with
absorbed elliptical host galaxies at $z=1-2$. It is likely that we are
observing some of the brighter members of the important new class of
X-ray Type II quasars.
\end{abstract}


\section{Introduction}
The X-ray Background (XRB) above 2\keV\ has at last been mostly
resolved into point sources by the {\sl Chandra} X-ray Observatory
(Mushotzky et al 2000; Brandt et al 2000). {\sl Chandra}'s superb
sub-arcsecond imaging provided the high-sensitivity confusion-less
images for this breakthrough.

The generally accepted model for the XRB is that it is dominated by
absorbed active galactic nuclei (AGN: Setti \& Woltjer 1989; Madau,
Ghisellini \& Fabian 1994; Comastri et al 1995), a significant
fraction of which are Compton-thick (i.e. have an absorbing column
$>10^{24}\pcmsq$): collectively such objects are Type~II AGN. So far,
only the 2-7 keV XRB has been (mostly) resolved with {\sl Chandra}.
This regime is not expected to be sensitive to large numbers of
Compton-thick objects, where the X-ray emission emerges only above
5\keV\ (Wilman \& Fabian 1999; Wilman, Fabian \& Nulsen 2000); it
should reveal sources with column densities more typically of
$10^{22-23}$\pcmsq. The Compton-thick objects provide much of the
power where the XRB spectrum peaks in $\nu I_{\nu}$. Correction of the
XRB for absorption (Fabian \& Iwasawa 1999) shows that accretion at
the standard 10 per cent efficiency onto massive black holes can
approximately account for the local mass density in black holes
(Magorrian et al 1998) and that about 85 per cent of that accretion
power is absorbed and re-radiated in the mid- to far-IR.

Despite the recent progress resolving the hard XRB into discrete
sources with the correct collective spectrum, the actual
identification of many of these objects is, however, not
straightforward. Roughly one-third are blue broad-line quasars,
another third are identified with faint, optically-normal galaxies and
the final third have only extremely faint optical counterparts or no
detectable counterpart at all (Mushotzky et al 2000; Brandt et al
2000; Maiolino et al 2000). Mushotzky et al (2000) find infrared HK$'$
band counterparts for most of their X-ray detected sources.

We have used SCUBA maps of the cores of the lensing clusters A2390 and
A1835 to place deep submillimetre limits on the three serendipitous
{\sl Chandra} sources which lie in the field (Fabian et al 2000). Only
one (marginal) source is detected in both the X-ray and submillimetre bands.
Of three X-ray sources in the HST field of A2390 we find that one
plausibly has a photometric redshift of 0.9 and conclude from its hard
X-ray spectrum that it is a Type~II quasar; the other two have $V>26$
(Fabian et al 2000). Such objects are therefore difficult to follow up
in the optical band alone. Although two SCUBA sources in the A370
field that optically resemble AGN are detected in X-rays (Bautz et al,
in preparation), none of the 10 SCUBA sources in fields flanking the
Hubble Deep Field is detected in a deep Chandra observation
(Hornschemeier et al 2000).

In this paper we present observations which are part of a programme to
determine the origin of the optically-faint, hard X-ray sources that
are most likely responsible for the hard X-ray background. Given the
possibility that the sources are highly redshifted and/or obscured, we
have sought infrared counterparts at the X-ray target position. Our
work bridges other surveys in this field which are either very deep,
or shallow with a wider area coverage. By targeting the serendipitous
sources from several 10-20~ksec {\sl Chandra} observations, we are
able to select the very brightest absorbed sources in each field for 
follow-up. 

\section{Selection and properties of X-ray targets }
The X-ray sources were selected from those found serendipitously in
the field of {\sl Chandra} observations of galaxy clusters obtained
during the Guaranteed Time of one of us (ACF). These observations were
typically of 10-20~ksec duration (see Table~1) and all were taken so
that the on-axis pointing position fell on the ACIS-S3 detector, apart
from that with sequence number 800010.  We shall present a detailed
analysis of the X-ray properties of the entire sample of serendipitous
sources elsewhere; here we detail follow-up observations of a
preliminary small sample of sources of a very specific type.

We examined each of the cluster fields for serendipitous sources using
the Chandra Interactive Analysis of Observations (CIAO) detection
software. All three available detect algorithms (celldetect, wavdetect
and vtpdetect) were used in order to search for sources under the
different assumptions and methods that each employ. In the
first instance, we ran the detect software on unbinned data in three
energy bands: 0.5-7\keV, \lq soft' 0.5-2\keV\ and \lq hard' 2-7\keV.
Those sources with detectable hard counts were then selected for more
detailed follow-up. As detailed later (section~\ref{sec:sohrats}), the
higher sensitivity of {\sl Chandra} below 2\keV\ means that even
genuinely hard sources can still show plenty of counts in our soft
band. Most of the sources in this paper were detected close to the
pointing position, either in the S3 or S2 chips. The exceptions were
the two sources in the field of A2199 (CXOU J162850.9+392434 and
CXOU J162827.8+392343), which were both in I3, and
CXOU J031946.4+413734 and CXOU J031946.2+413737 in the Perseus field
which were in the I2 chip. The sources were 
2-9 arcminutes off-axis, with a reduced effective
area down to 80 per cent of the on-axis value.

We calculated the number of counts in each of these three energy
bands using a box typically of side 16 pixels (8 arcsec; although the
box size was increased for sources well off-axis, ie for
CXOU J162827.8+392343, CXOU J091340.9+410314 and
CXOU J140048.4+024954) from which to extract the source counts. The
background countrate was estimated from a box around the source
(excluding the source box) of sidelength typically 55 pixels (27.5
arcsec; or larger for those objects with countrate extracted from a
larger source box). The only exception was in the case of
CXOU J031946.4+413734 and CXOU J031946.2+413737 in the Perseus field,
where the sources were so close that the background rate for both was
estimated from a close region of sky. The counts in all three energy
bands, as well as the ratio of soft/hard counts are given in Table~1.

As the {\sl Chandra} observations were relatively early on in the
operation of the satellite, all were affected by inaccuracies in the
aspect solution during the initial pipeline processing of the data.
This led to offsets of up to 8 arcseconds between the coordinate
positions given by {\sl Chandra} and those of the sky. 
%                x    y
%xrays perseus -2.40 -0.41 
%xrayi perseus -1.86 -0.88
%IRAS09        +1.92 +0.53
%A1795         +7.69 -3.12
%A1835         +9.07 -3.23
%A2199         +3.96 -8.32
This offset was easy to correct for in the observations of
IRAS~09104+4109 (800017) and Perseus (800010), where the
position of the AGN can be determined with half-arcsecond precision
from the {\sl Chandra} hard-band image, to be compared to
sub-arcsecond radio core position. A hard nuclear peak was not, however,
found in the {\sl Chandra} images of A1795, A1835 or A2199, so the
X-ray optical registration was done by a systematic cross-correlation
between {\sl all} the X-ray sources against optical sources from the
Digitized Sky Survey (DSS). Through this process we improved the
accuracy of the attitude solution to within about 2 arcseconds. Note
we assume that only a simple translation in sky coordinates is
required, with no rotation or stretching. The (registered) X-ray
source positions were then compared to optical sources on the DSS.
Although we have been using multi-colour (archival) optical imaging
data of our fields, many of the sources lie a few arcmin out from the
centre of the cluster, and are not always covered by previous
observations.

This inaccurate aspect solution to the X-ray data and the
uncertainties inherent in the correction of the {\sl Chandra}
coordinates to the sky presents some confusion when applied to
conventional naming of the source. The coordinates used to name the
source are derived from a definite ID of the source from IR imaging
presented later in this paper. Where no source is detected in the IR
imaging we use the IR-detected sources within this field to apply the
correct offset between the {\sl Chandra} frame of reference and the
sky, and thus derive the source name.

We selected our targets from those sources that had a soft-to-hard (S/H) ratio of
less than 3.5 (see Table~1; also section~\ref{sec:sohrats} for
implications of the S/H ratio), and either a very faint, or no, optical
identification on the DSS. The exceptions were two intriguing sources
in the ACIS-I observation of the Perseus cluster. Their unusually
close proximity to each other (5~arcsec) and exact association with two close
optical sources marked them out as particularly interesting and worthy
of detailed follow-up. The DSS B- and R-band images around each source
observed in this paper are shown in Fig~1, along with matching {\sl
Chandra} images in the 0.5-7\keV, 0.5-2\keV\ and 2-7\keV\ energy
bands. Faint optical identifications are visible not only for
CXOU J031946.4+413734 and CXOU J031946.2+413737, but also
 CXOU J091357.5+405938, CXOU J162827.8+392343 and (perhaps)
CXOU J091360.0+405548. The faint optical source seen in the
CXOU J134905.8+263752 box is several arcseconds away from the X-ray
position, and thus we assume it is not associated.

\section{Observations}

\subsection{Optical spectra}
Optical spectra of the close X-ray sources CXOU J031946.4+413734 and
CXOU J031946.2+413737 were obtained in service time with the ISIS
double beam spectrograph on the William Herschel Telescope (WHT) on La
Palma, during the night of 1999 Dec 15. The 1 arcsec-wide slit was
oriented at a position angle of 138\degmark\ in order to detect the
emission from both sources. The total exposure was 3000\s, and the
R158R and R158B gratings were used to produce 
a wavelength range of 3500-5500\AA\ on the EEV chip
in the blue arm, and 5150-8050\AA\ on the Tek chip in the red arm of
the instrument. The data were bias-subtracted, flat-fielded,
wavelength-calibrated from exposures of an arclamp, and corrected for
the Galactic extinction of E(B-V)=0.31 in this direction.

\subsection{Infrared observations}
Near-infrared spectra and images were taken during the nights of 2000
Feb 24-25 at the United Kingdom Infrared Telescope (UKIRT) in
Hawaii. The night of Feb 24 was of very good seeing and transparency,
both of which, however, deteriorated by the time of our observations
on Feb 25. We used two instruments: the 2D grating spectrometer CGS4,
and the cooled infrared camera IRCAM3/TUFTI. A full log of
observations is shown in Table~\ref{tab:irlog}. 

CGS4 was used with the 40~l/mm grating on the long camera, yielding a
pixel scale of 0.618 arcsec per pixel. The slit width was set to 4
pixels (ie 2.47 arcsec), and observations were taken using the
standard \lq quad-slide' nodding pattern of a-b-b-a along the slit.
Spectra were taken in the first order in each of the H and K bands on
some objects, as well as of corresponding spectrophotometric and
atmospheric absorption standards in the same band. The data were
reduced using the standard reduction package CGS4DR V1.3-0. The object
spectrum was divided by the spectrum of a standard star observed at
approximately the air mass, assumed to
approximate to a black body, in order to remove any common atmospheric
features. The object spectra were flux-calibrated using a standard
star; the flux calibration  was estimated to be accurate to within
ten per cent. The final, calibrated spectrum of the object was extracted from three
rows centred on the object peakup row, and three rows around the sky
peakup row 19 pixels away.

IRCAM3/TUFTI is an imaging camera with a scale of 0.0814 arcsec/pixel,
and a total field of view of 20.8$\times$20.8 arcsec. TUFTI was
used in ND~STARE mode, with standard read-out. Objects were observed
using a jitter grid of 9 points, each separated by 6 arcsec.
We observed for 60~s at each of
the grid points, except for CXOU J140100.2+025720, leading to total exposure time of 540~s in each of
the J,H and K bands. The imaging observations for
CXOU J140100.2+025720 were curtailed because we were approaching
the end of that night's observing. The data were reduced using the
standard ORACDR reduction package available from UKIRT, and
flux-calibrated using observations of several standard stars in the
same bands.

%These fluxes can also be compared
%to fluxes present in the images taken from IRCAM3/TUFTI, whenever both
%spectra and images are available for the same object. All
%corresponding fluxes are within a factor of less than 2.5 of each
%other.

\section{Results and Discussion }
\subsection{X-ray luminosities}
We derive the bolometric luminosity of the X-ray sources from the
observed 0.5-7\keV\ count rate, assuming the emission originates in a
non-thermal power-law with photon index of $\Gamma=2$, subject only to
Galactic absorption. The observed 0.5-7\keV\ fluxes range over
$4.3-63.3\times10^{-15}$\ergpcmsqps (Table~1), implying a bolometric
luminosity range of $1.7-24.8\times10^{44}$ and
$9.0-130.8\times10^{45}$\ergps for redshifts of 0.5 and 3
respectively.

\subsection{Optical spectra}
The optical spectra of CXOU J031946.4+413734 and CXOU J031946.2+413737
are shown in Fig~\ref{fig:cc1optsp}. CXOU J031946.4+413734 (the
optically-fainter, X-ray-harder source to the south-east) has a
solitary broad line observed at 6454\AA, with a FWHM of 5460\kmps and
an intensity of 1.6$\times10^{-15}$\ergpspcmsq. The most likely
identification for this is MgII, implying a redshift for the source of
$z=1.307$; however we do not obviously see CII]$\lambda$2336 or
CIII]$\lambda$1909 emission line at 5365\AA\ and 4404\AA, or H$\alpha$
at 1.514$\mu$m in the infrared spectrum (Fig~\ref{fig:irsp}; the
2-$\sigma$ emission line nearby is at 1.504$\mu$m). The line can only
be CIII] if the expected Ly$\alpha$ at 4110\AA\ is completely
extinguished by dust absorption; it is unlikely to be Ly$\alpha$ itself, as the continuum
does not cut off due to a Ly$\alpha$-forest as expected for a source
at $z=4.3$. CXOU J031946.2+413737 shows a relatively featureless
continuum, with too little flux for a reliable redshift to be obtained
from cross-correlation with a template galaxy spectrum.

\subsection{Infrared spectra}
We detected continuum emission from four of the objects observed with
CGS4, only failing to detect any signal at the position of
CXOU J091360.0+405548 at an upper limit of
$5.1\times10^{-19}$\ergpspcmsqpA. (CXOU J091360.0+405548 is the
source we also failed to detect from the infrared imaging, see next
section). All the detected objects have flat infrared spectra with no
significant emission-line features (Fig~\ref{fig:irsp}). Where
magnitudes from the infrared imaging are available (see next
section), they are compatible with the spectra (see
Fig~\ref{fig:irsp} for a direct comparison). The flux inferred from
the infrared imaging tends to be higher than that determined from the
spectra, which is to be expected if our slit placement does not quite
cover all the light from the object.

We searched for emission lines in those parts of the infrared spectrum
least affected by any OH atmospheric absorption (if the sky is
variable then we may not be able to correct for these features fully
in the data reduction and spurious emission lines would be introduced;
these regions of the spectrum are shown by the dotted lines in
Fig~\ref{fig:irsp}). We find no significant emission lines, with 3-$\sigma$ upper
limits to the equivalent width of 185-720\AA\ in H, and
145-190\AA\ in K (depending on the length of the exposure). We would
expect some Balmer or Paschen line emission to be visible for targets
in most of the redshift range $0<z<4$, and MgII to be visible for
sources $4.3<z<5.4$. It is thus surprising that none
of our targets show strong emission lines in the CGS4 spectra. This
suggests that any active nucleus -- and any ionized gas that surrounds
it -- is heavily obscured.

\subsection{Infrared imaging}
Of the nine sources imaged in good seeing conditions, we detect
 infrared sources in the 21$\times$21 arcsec TUFTI field of view for
 eight of them. A further three objects attempted during conditions of
 poor seeing and transparency were not detected. Due to the
 variability of the conditions we are unable to quantify the detection
 limits. CXOU J091360.0+405548 (our least significant X-ray source)
 was not detected to a 3-$\sigma$ limit of 22.66 in J, 22.27 in H and
21.62 in K (assuming a point source). The J, H and K-band
 TUFTI images are shown in Fig~4 for all sources
 where infrared sources were found in the field. In
 Table~\ref{tab:irmags} we list the infrared magnitudes (and estimated
 DSS B and R magnitudes or limits) of each source near the centre of
 the TUFTI image, along with its offset from the boresight of TUFTI
 (and thus from the X-ray position). The position of this aim-point
 and its constancy was established from the observations of standard
 stars throughout the night.

There is a very clear single identification of a source near the
centre of the TUFTI field of view for seven of the eight sources shown
in Fig~4. The source for CXOU J140106.9+024934 is
resolved into three separate components in the K band. CXOU
J091352.8+405829 is the only source (out of those closest to the
centre) to show any significant ellipticity. The registration of the
X-ray coordinates to the sky (as indicated by the offset of the
source from the aimpoint of TUFTI) is, not surprisingly, best in the
field of IRAS~09104+4109, the source with a clear detection of the
point-like AGN at hard energies to cross-correlate with the radio
position. Offsets are larger for the other fields, but all are good to
within 2.1 arcsec. Only CXOU J140100.2+025720 has no clear
identification, as there are two sources that are both at a larger
offset from the centre than expected: it is not clear that either of
these are the identification of the X-ray source. For clarity we will,
however, refer to them as CXOU J140100.2+025720 (SSE) and (ESE),
according to the direction of the offset from the position of the
X-ray source CXOU J140100.2+025720. There is no correlation between
the J-K colours of our detected objects and the ratio of their soft to
hard X-ray flux, but this is not unexpected as we have deliberately
chosen a very small range of X-ray flux ratio to investigate in the
first place.

\subsection{Absorbed AGN?}
\label{sec:sohrats}
We consider it unlikely that our detected objects are optically-faint
stars, as any star with a sufficient level of X-ray activity is
unlikely to show such a featureless infrared continuum. The infrared
colours of our detected targets are very different to those of main
sequence, giant and supergiant stars of all spectral types
(Fig~\ref{fig:ircols}), although the infrared colours alone are
consistent with some IR-selected stars (eg the samples shown in Fig~5
of Dickinson et al 2000). We note also that stars comprised only 6 per
cent of the sources found in the ROSAT Deep Survey (Schmidt et al
1998). The infrared colours of our sources are better matched to
those of X-ray-selected quasars in the compilation of Elvis et al
(1994).

%(Errors are 1 sigma)

We performed a simple X-ray spectral analysis of the two brightest
sources, CXOU J134905.8+263752 and CXOU J134849.0+263716 using XSPEC.
The spectra were fitted with power-law models of fixed photon index
$\Gamma=2$, absorbed (in our frame) by a column density N$_{\rm H}$.
The X-ray spectra require excess column densities of N$_{\rm
H}=2.6\pm0.7$ and $3.5^{+1.3}_{-1.1}\times10^{21}$\pcmsq, for CXOU
J134905.8+263752 and CXOU J134849.0+263716 (errors are 1-$\sigma$).
For example, the fit to CXOU J134905.8+263752 for a power-law model
(of fixed $\Gamma$=2) with only the Galactic absorption gives
$\chi^2$=24.5 (for 12 degrees of freedom), improving to $\chi^2$=6.1
(for 11 d.o.f.) when the absorption is allowed to vary freely, giving
N$_{\rm H}=2.6\pm0.7\times10^{21}$\pcmsq (Fig~\ref{fig:ce16xrays}). A
fit with the absorption fixed at Galactic, and now with the power-law
slope varying freely yields $\Gamma=1.4\pm0.14$, with $\chi^2$=9.6 (11
d.o.f.), and a completely free fit gives $\Gamma=1.9\pm0.2$ and
N$_{\rm H}=2.3\pm0.9\times10^{21}$\pcmsq, with $\chi^2$=6.06 (for 10
d.o.f.). Thus there is a greater than 95 per cent probability that the
source requires excess absorption over the Galactic column. The models
with $\Gamma=2$ and free N$_{\rm H}$ predict de-absorbed fluxes at
1keV of $2.0\pm0.3\times10^{-5}$ and $8.8^{+1.8}_{-1.7}\times10^{-6}$
photons\pcmsqpkeVps for CXOU J134905.8+263752 and CXOU
J134849.0+263716 respectively (equivalent to 13 and 6~nJy).

We converted these de-absorbed fluxes at 1\keV\ to intrinsic K-band
magnitudes, assuming a spectral index of $\alpha_{\rm KX}=1.16$ (such
that $S\propto\nu^{-\alpha}$). This index is the median for the sample
of 41 quasars in Elvis et al (1994), who chose optically-bright
objects with high signal-to-noise ratio {\sl Einstein} X-ray spectra.
They are predominantly low redshift objects, so K-corrections should
be negligible. The predicted K-band magnitude of 17.3 for CXOU
J134905.8+263752 is slightly fainter than the observed value of
17.01$\pm$0.06, whereas the predicted magnitude of 18.2 for CXOU
J134849.0+263716 is a lot brighter than the observed value of
19.99$\pm$0.25. This suggests that for the total K-band light of CXOU
J134849.0+263716 to be due wholly to the AGN continuum, it must be
reddened by $\sim2$ magnitudes at K. For a Galactic dust-to-gas ratio,
this equates to N$_{\rm H}=3.5\times10^{22}\pcmsq$, substantially more
than the value derived from the X-ray fitting.

%[Note that for an intrinsic $\Gamma=2$ power-law
%X-ray spectrum, there is no K-correction associated with converting a
%de-absorbed monochromatic flux at E$_{\rm obs}=1$\keV\ to one at
%E$_{\rm rest}=1$\keV; for very high redshift objects ($z>3.4$),
%however, the observed K-band will begin to sample the optical/UV bump
%which is likely to deviate from the $\alpha=1.22$ power-law assumed.
%In such cases, the estimated K-band extinctions to the nucleus will
%underestimate the true values.]

We extend this calculation to predict optical and infrared magnitudes
for all our TUFTI-detected sources, assuming that all the emission is
due to a typical quasar at a range of redshifts. We use PIMMS to
obtain the normalization of the quasar spectral energy distribution
(SED) at 1~keV from the observed 0.5-7\keV\ count rate, assuming
$\Gamma=2$ using the Galactic column. We
then extrapolate from the inferred X-ray flux at 2\keV\ using the
(rest-frame) SED of a typical radio-quiet quasar, approximated from
that given in Elvis et al (1994); $\alpha_{OX}=1.4$ (bridging the
monochromatic luminosities at 2500\AA\ and 2keV), $\alpha=0.15$ (over
0.25-1.28$\mu$m), $\alpha=2.38$ (1.28-1.85$\mu$m) and $\alpha=1.38$
($>$1.85$\mu$m).

  For all the sources, the observed SEDs are redder in
slope than that expected from the quasar predictions
(Fig~\ref{fig:seds}).  The majority of the objects have unabsorbed
quasar predictions that are insufficient (by up to 1.5 magnitudes) to
account for the observed magnitudes, suggesting that the AGN continuum
is not the main contributor to the infrared magnitudes. Where the
(low-redshift) quasar SEDs are more compatible with the observed
infrared magnitudes (CXOU J091340.9+410314, CXOU J134905.8+263752) the
quasar SED still overestimates the optical band limits. There remains
only one source (CXOU J134849.0+263716) where the extrapolated
spectrum vastly overestimates the observed magnitudes; here any AGN
component must be reddened by at least 1.5 magnitudes in the infrared.
These results are, however, sensitive to the $\alpha_{OX}$ used:
varying $\alpha_{OX}$ by only $\pm0.1$ leads to a corresponding
$\pm0.6$ mag variation in the optical-infrared magnitudes shown in
Fig~\ref{fig:seds}. Even if $\alpha_{OX}$ is adjusted so that the
quasar can account for all of the K-band light, its continuum must be
highly reddened to also fit the observed B and R limits or magnitudes
(we assume that the lack of detection in the optical is not due to
variability). The results indicate both that a host galaxy contributes
much of the infrared light of many of our sources, and that any
contribution from the quasar continuum requires significant reddening.

Further constraints on the redshift and N$_{\rm H}$ may be obtained by
comparing the soft to hard (S/H) count ratios (Table~1) with those
predicted by XSPEC models, as shown in Table~\ref{tab:sohrats} and
Fig~\ref{fig:sohrats}. For a given model (ie $z$ and N$_{\rm H}$), the
differences between the predictions for the two brightest sources reflect the
different response matrices of the front- (CXOU J134905.8+263752 on
S2) and back- (CXOU J134849.0+263716 on S3) illuminated CCDs. 
%This
%discrepancy is most apparent for the hardest models (ie high
%absorption at low redshift), where the S/H ratios are much higher for
%the front-illuminated chips. At first sight this appears to contradict
%the fact that the back-illuminated CCDs are the more sensitive of the
%two to soft photons. It is, however, due to the degradation in the
%energy resolution of the front-illuminated chips (caused by low-energy
%protons early in the mission; Townsley et al 2000) -- for such hard
%models, essentially all of the counts appearing in the soft-band for
%the front-illuminated CCD are due to the tail of the response to
%incident photons of much higher energy. Thus even a hard source can
%appear as a \lq soft'.
The observed S/H ratios are $2.4\pm0.4$ (for CXOU J134905.8+263752)
and $2.5\pm0.5$ (for CXOU J134849.0+263716). Very roughly, the values
in Table~\ref{tab:sohrats} suggest that if our two sources
had power-law spectra, they have N$_{\rm H}\sim10^{22}$\pcmsq if at
low redshift, or more like N$_{\rm H}\sim10^{23}$\pcmsq if at $z\ge1$,
As a first approximation, the Tables may also be used to interpret the
S/H count ratios for the other sources, for which there are
insufficient counts to justify spectral fitting [using the CXOU
J134905.8+263752 (and CXOU J134849.0+263716) predicted values as
guides for sources on the front- (back-)illuminated chips]. We have
attempted simple spectral fitting of the sources with around 50
counts, and although the error bars are larger, the results confirm
the conclusions of Table~\ref{tab:sohrats}. The X-ray colours of all
our sources are consistent with the X-ray emission originating in an
AGN at a range of redshift, but {\sl only} if that emission is
absorbed by an intrinsic column density of at least N$_{\rm
H}\sim10^{22}$\pcmsq at $z>0.5$. We conclude that the bulk of our
objects have X-ray column densities which classify them as
Compton-thin Type~II objects.

\twocolumn
\subsection{Photometric redshifts}
Given that at least one of the detected infrared objects is resolved
to have significant ellipticity, and that the host galaxies of AGN
usually make the dominant contribution to the near-infrared light of
AGN (Rix et al 1999), we also compare the infrared colours and
magnitudes to those of galaxies at a range of redshift.  The lack of
strong emission lines in the infrared spectra argues against strong
starburst behaviour, so we consider only elliptical galaxies. We used
the programme HYPERZ (Bolzonella, Pello \& Miralles 2000), which fits
a grid of template galaxy spectra (generated from the GISSEL library;
Bruzual \& Charlot 1993) to the observed magnitudes, with variations
permitted in the age, redshift and intrinsic reddening. We attempted a
variety of fits: a 3~Gyr-old elliptical galaxy (with e-folding time
$\tau$ of 1~Gyr) fitted without and with the possibility of intrinsic
reddening; and a choice between an elliptical galaxy at ages 1.5, 3
and 5~Gyr again without and then with reddening. In all cases we
assumed the reddening law of Calzetti et al (2000) and constrained the
age so that it did not exceed the age of the Universe at the redshift
under consideration (assuming $H_0=50$\kmpspMpc). The
maximum redshift considered was $z=6$.

Our results for the basic model (a 3~Gyr-old elliptical with no
intrinsic reddening) are shown for three of our detected sources in
Table~\ref{tab:photzfit}. We also plot the best-fit solutions for this
model to CXOU J134905.8+263752, CXOU J134849.0+263716 and CXOU
J140106.9+024934 in Fig~\ref{fig:photzfit}. Increasing the range of
possible elliptical galaxy ages, and/or allowing the possibility of
intrinsic reddening to this model decreased the $\chi^2$, but
broadened the range of possible redshifts, and sometimes introduced a
secondary $\chi^2$ minimum at lower redshift. Without deeper optical
limits it is hard to constrain the redshift in any detail;
we can only say that the infrared colours are not inconsistent with
an origin from elliptical galaxies above a redshift of one. 

Host galaxies to radio-quiet quasars generally have a modest
luminosity of just less than, or around that of an L$^*$ galaxy, with
radio-loud quasars lying in galaxies a factor of 2-5 times brighter
(Rix et al 1999). We thus also compare our observed infrared
magnitudes to those expected from a passively-evolving, unabsorbed
L$^*$ elliptical galaxy (R. McMahon \& A. Aragon-Salamanca, private
communication) in Fig~\ref{fig:magcomp}. This crude comparison of
magnitudes suggests that CXOU J140106.9+024934, CXOU J134849.0+263716
and CXOU J140100.2+025720 (ESE) are consistent with L$^*$ galaxies at a
redshift of $z=2$, with other sources such as CXOU J140100.2+025720
(SSE), CXOU J091340.9+410314, CXOU J091352.8+405829 and CXOU
J162827.8+392343 being nearer $z=1$. The inferred redshifts may be
underestimated, however, if the infrared colours and magnitudes also
include a contribution from a central quasar continuum.

\section{Summary and conclusions}

We have carried out follow-up observations of optically-faint, X-ray
hard, serendipitous {\sl Chandra} sources, and find that they are
readily detected in the near-infrared. Spectra in the infrared of some
of them appear flat and featureless, suggesting that strong
emission-line activity is either absent or heavily obscured. Only one
source -- which is the brightest optically -- shows a strong emission
line in an optical spectrum, which we cannot identify unambiguously.
The 0.5-7\keV\ fluxes of our sources are in the range
$0.3-4.6\times10^{-14}$\ergpcmsqps which, if due to an unabsorbed
non-thermal quasar spectrum, imply bolometric luminosities of a few
times $10^{44}-10^{46}$\ergps, for redshifts between $0.1<z<3$. The
X-ray colours of all our sources are consistent with an origin in
quasars only if absorbed by an intrinsic column density of at least
$10^{22}$\pcmsq. If we extrapolate a typical radio-quiet quasar
spectrum at $0<z<3$ from the observed X-ray flux into the optical and
infrared wavebands, we find that a host galaxy probably contributes
much of the infrared light, and that the optical quasar continuum
requires significant reddening. The infrared magnitudes and colours
are consistent with a (host) L$\sim$L$^*$galaxy at moderate redshifts
$0.5<z<2.5$.

Although we do not have deep optical images for most of our
objects, the limits imply that the optical and infrared properties may
resemble so-called extremely red objects (EROs: see Scodeggio \& Silva
2000 and references therein). These have proven difficult to follow up
even with large telescopes, and good spectra and redshifts are only
available for a small fraction. Whether there is a deeper, physical
connection between optically faint Chandra sources and EROs must await
larger samples of Chandra sources with deep optical and infrared
coverage.

The properties of our small sample of objects is consistent with the
existence of a population of moderately absorbed (i.e. Compton thin
Type II) quasars at $z=1-2$, as predicted by recent models for the
hard XRB (Madau et al 1994; Celotti et al 1995; Comastri et al 1995;
Wilman \& Fabian 1999). 

Finally, we note that the objects we have likely found, namely Type II
quasars, are qualitatively different from Type II Seyferts. They do
not obviously have any strong narrow-line region (although deeper
spectra covering a wider band on more objects are needed to be
definite on this issue), nor any obvious scattered blue continuum or
blue light from star formation. They would not even be classified as
active galaxies on the basis of what we have seen so far of their
optical and infrared properties. It is only on the basis of the X-ray
emission that we classify them as Type II objects, where Type II means
strongly obscured (see also the discussion in Matt et al 2000). An
appropriate name is X-ray Type~II quasar.  Of course, if 85 per cent
of accretion power is absorbed (Fabian \& Iwasawa 1999) then the
objects dominating that power cannot have the broad torus
opening-angles commonly ascribed to Seyfert galaxies. The simple and
successful geometrical unification scheme for Seyfert galaxies (e.g.
Antonucci 1993) cannot extend to quasars.

\section{Acknowledgements}
We are grateful to the {\sl Chandra} project for the X-ray data, and to
M. Bolzonella, R. Pello and J.-M. Miralles for making HYPERZ
available. CSC and ACF thank the Royal Society, PG thanks the Isaac
Newton Trust and the Overseas Research Trust, and RJW and RMJ thank the
PPARC for financial support. We thank the service queue scheme on the
William Herschel Telescope, which is operated on the island of La
Palma by the Isaac Newton Group in the Spanish Observatorio del Roque
de los Muchachos of the Instituto de Astrofisica de Canarias. The
United Kingdom Infrared Telescope is operated by the Joint Astronomy
Centre on behalf of the U.K. Particle Physics and Astronomy Research
Council. This research has made use of the NASA/IPAC Extragalactic
Database (NED), and the Digitized Sky Surveys which were produced at
the Space Telescope Science Institute under U.S. Government grant NAG
W-2166.

{}

\begin{figure}
%\vspace{-2.0cm}
%\psfig{figure=fig1a.ps,width=1.1\textwidth,angle=180}
\psfig{figure=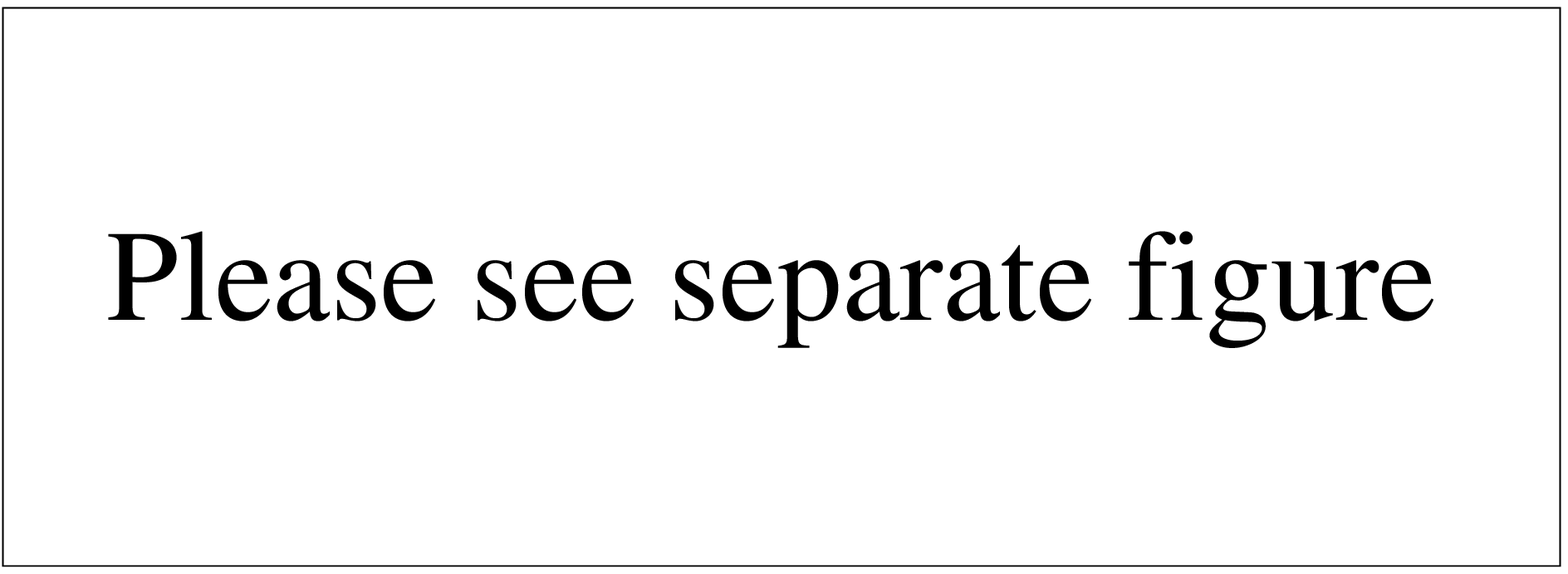,width=0.45\textwidth,angle=0}
\caption{Blue and red DSS images (40 arcsec on a side)
of the serendipitous {\em Chandra} sources. 
 Next (from left to right) are presented the 0.5-7\keV, the soft
(0.5-2\keV) and hard 
(2-7\keV) band {\em Chandra} images on   the same scale.
Where a box is drawn on an optical image it marks for comparison the
central
13$\times13$ arcmin of the TUFTI field shown in Fig~4. }
\end{figure}

%\addtocounter{figure}{+1}
\begin{figure*}
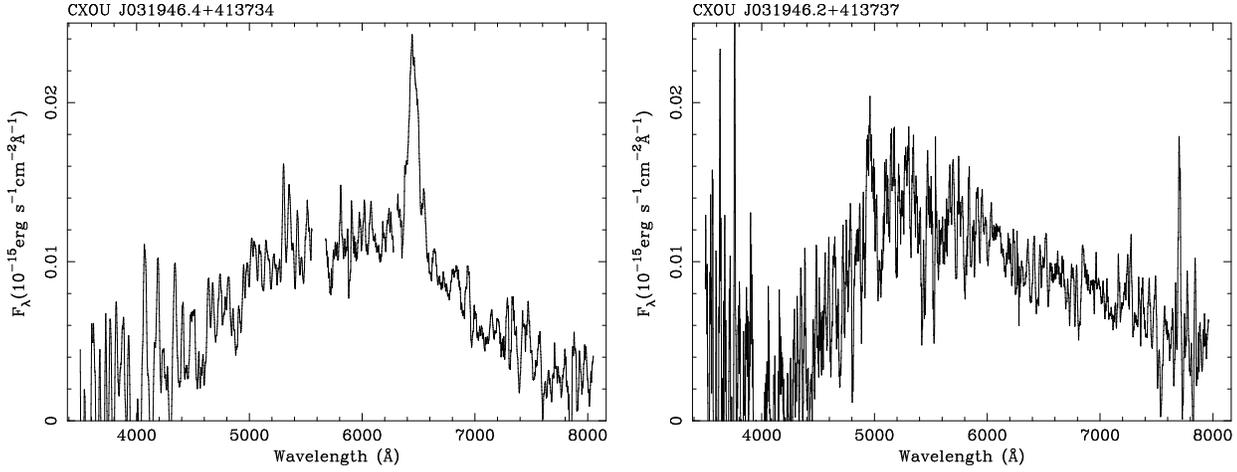

\hbox{
\psfig{figure=fig2l.ps,width=0.45\textwidth,angle=270}
\hspace{0.2cm}
\psfig{figure=fig2r.ps,width=0.45\textwidth,angle=270}}
\caption{\label{fig:cc1optsp} Optical spectra of the sources associated
with CXOU J031946.4+413734 (left) and CXOU J031946.2+413737 (right). The data have been
smoothed, and the spectra from the blue and red arms of ISIS have been
spliced at 5600\AA. } 
\end{figure*}

\begin{figure*}
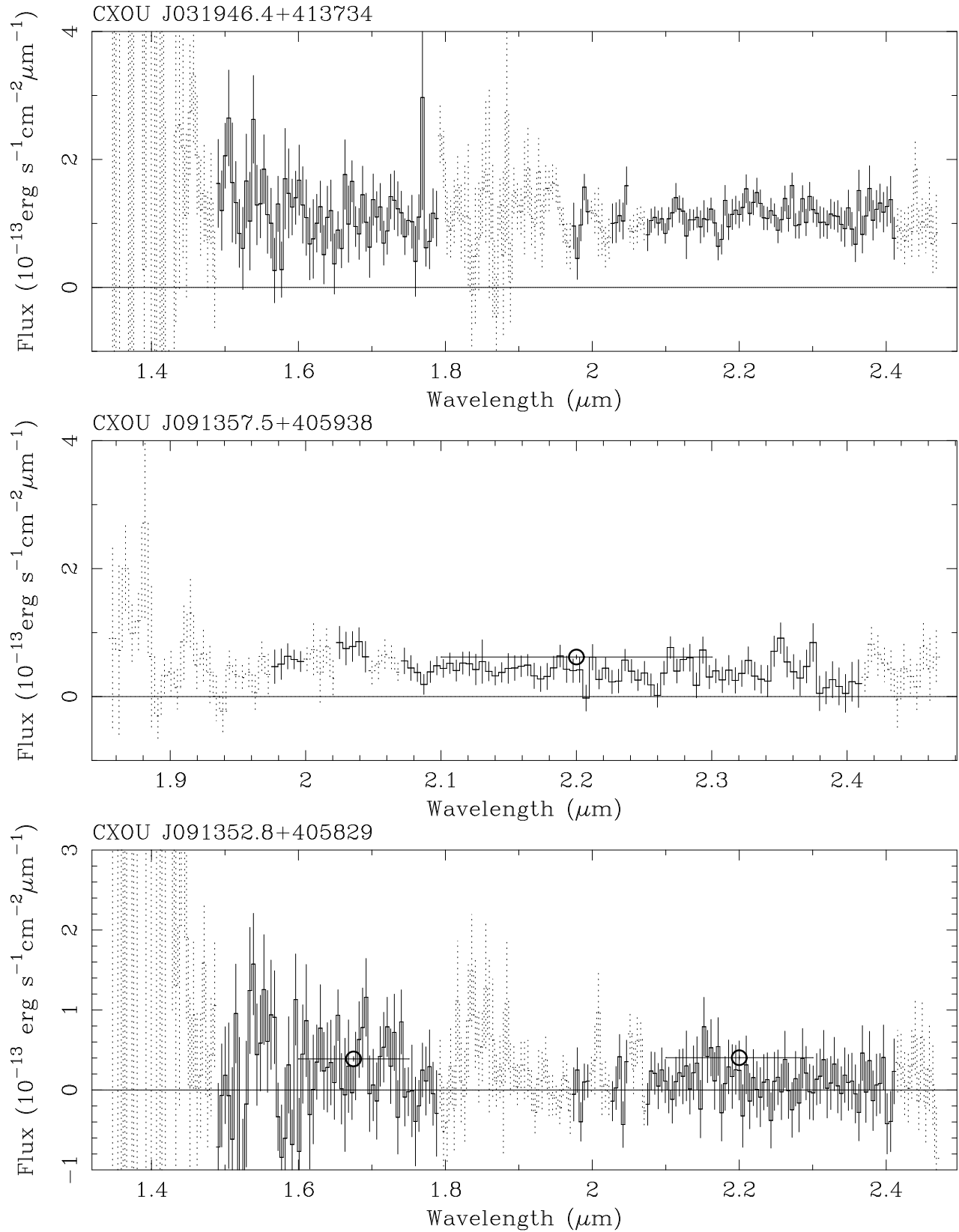

\vbox{
\psfig{figure=fig3a.ps,width=0.95\textwidth,angle=270}
\psfig{figure=fig3b.ps,width=0.95\textwidth,angle=270}
\psfig{figure=fig3c.ps,width=0.95\textwidth,angle=270}
}\caption{\label{fig:irsp} Infrared spectra of the sources associated
with (from top) CXOU J031946.4+413734, CXOU J091357.5+405938 and 
CXOU J091352.8+405829. The spectra have been binned by two, and regions where
the spectral features may be uncertain due to the sky
absorption spectrum are plotted in a dotted line. 
The open circles indicate the flux level determined from the H
and K band imaging, where available. 
}
\end{figure*}
\addtocounter{figure}{-1}
\begin{figure*}
\psfig{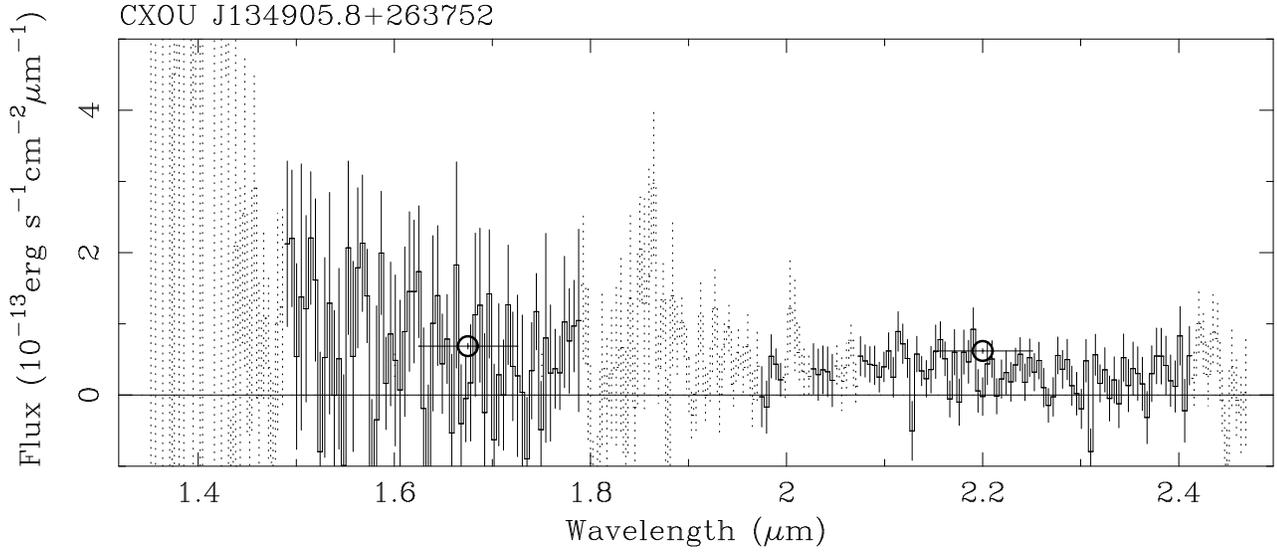}
\caption{\label{fig:irsp} (ctd.) Infrared spectrum of the source associated
with CXOU J134905.8+263752.}
\end{figure*}

\begin{figure}
%\vspace{-2cm}
%\psfig{figure=fig4a.ps,width=1.1\textwidth,angle=0}
\psfig{figure=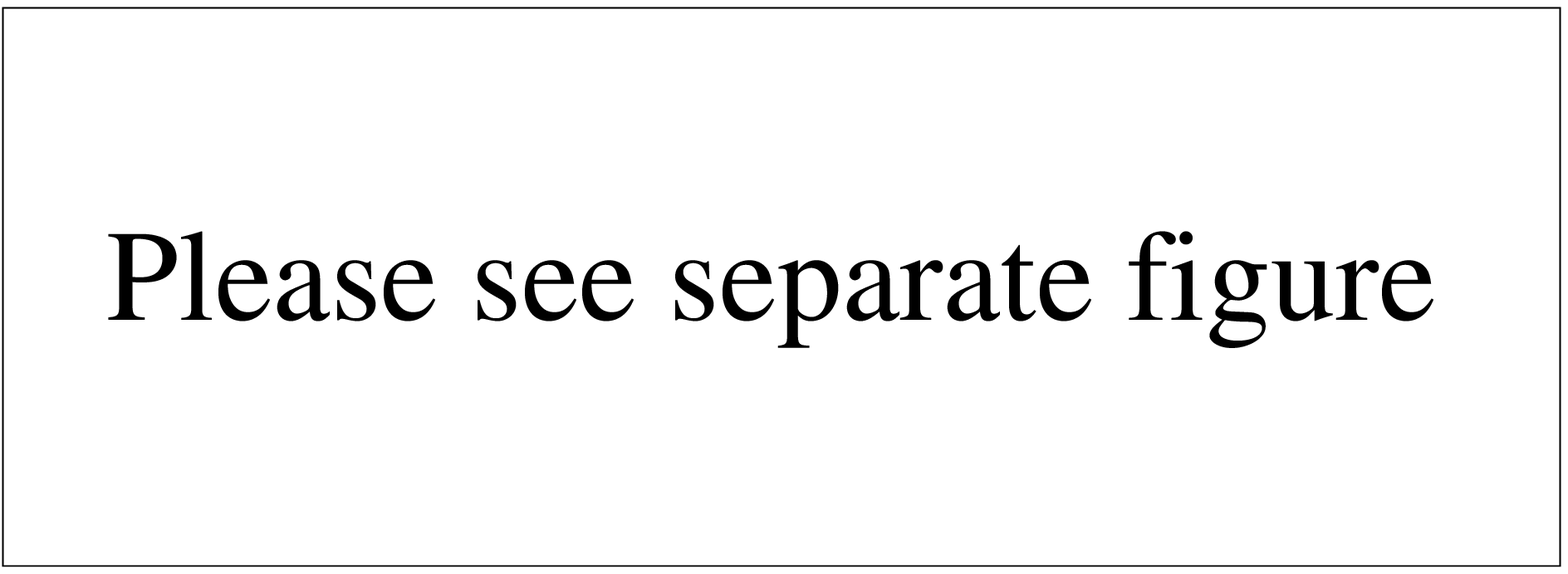,width=0.45\textwidth,angle=0}
\caption{Infra-red images of the central 13$\times$13
arcsec TUFTI field of view of the {\sl Chandra} sources. 
North is to the top, and the small
cross marks the position of the TUFTI boresight.}
%\label{fig:irimages}
\end{figure}
%\begin{figure*}
%\vspace{-2cm}
%\psfig{figure=fig4b.ps,width=1.1\textwidth,angle=0}
%\end{figure*}
%\twocolumn

%\addtocounter{figure}{+1}
\begin{figure*}
\psfig{figure=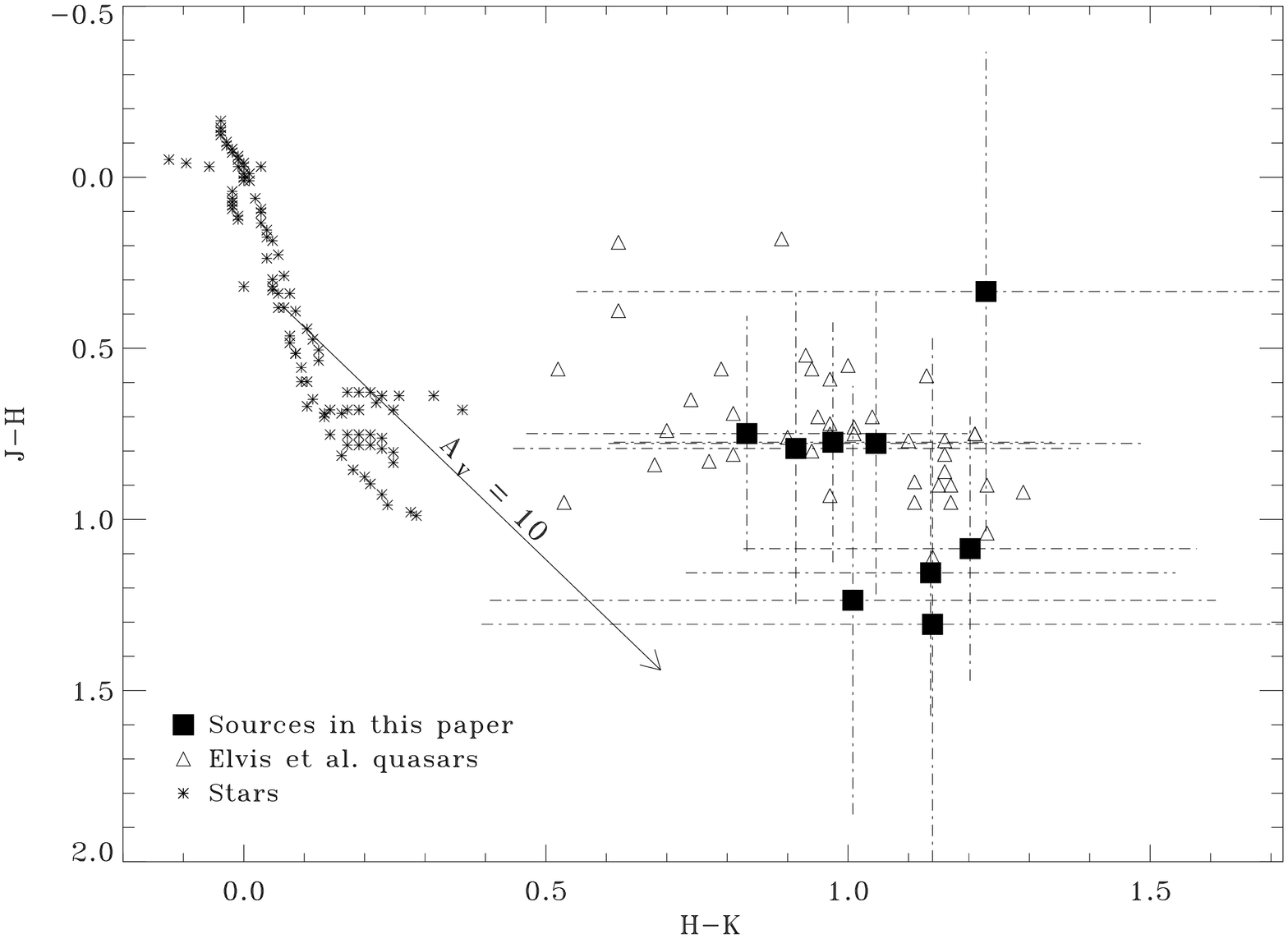,width=0.95\textwidth,angle=90}
\caption{\label{fig:ircols} 
Infrared colours of our detected sources (filled squares) plotted with
1-$\sigma$ errors  (dash-dot lines). The star symbols show the infrared
colours of main sequence, giant and supergiant stars of all spectral
types. The triangles show the colours of the X-ray selected quasars from
Elvis et al (1994; see text).   }
\end{figure*}

\begin{figure}
\psfig{figure=fig6.ps,width=0.45\textwidth,angle=270}
\caption{\label{fig:ce16xrays}
The X-ray spectrum of CXOU J134905.8+263752 (solid circle markers)
together with best fitting $\Gamma=2$ power-law model with free
absorption (solid line). The effect of that model of changing the
absorption to the Galactic value is indicated by the dotted line. The
{\sl x}-errors show the nominal bin width, whereas the {\sl y}-errors
are 1-$\sigma$. }
\end{figure}

\begin{figure*}
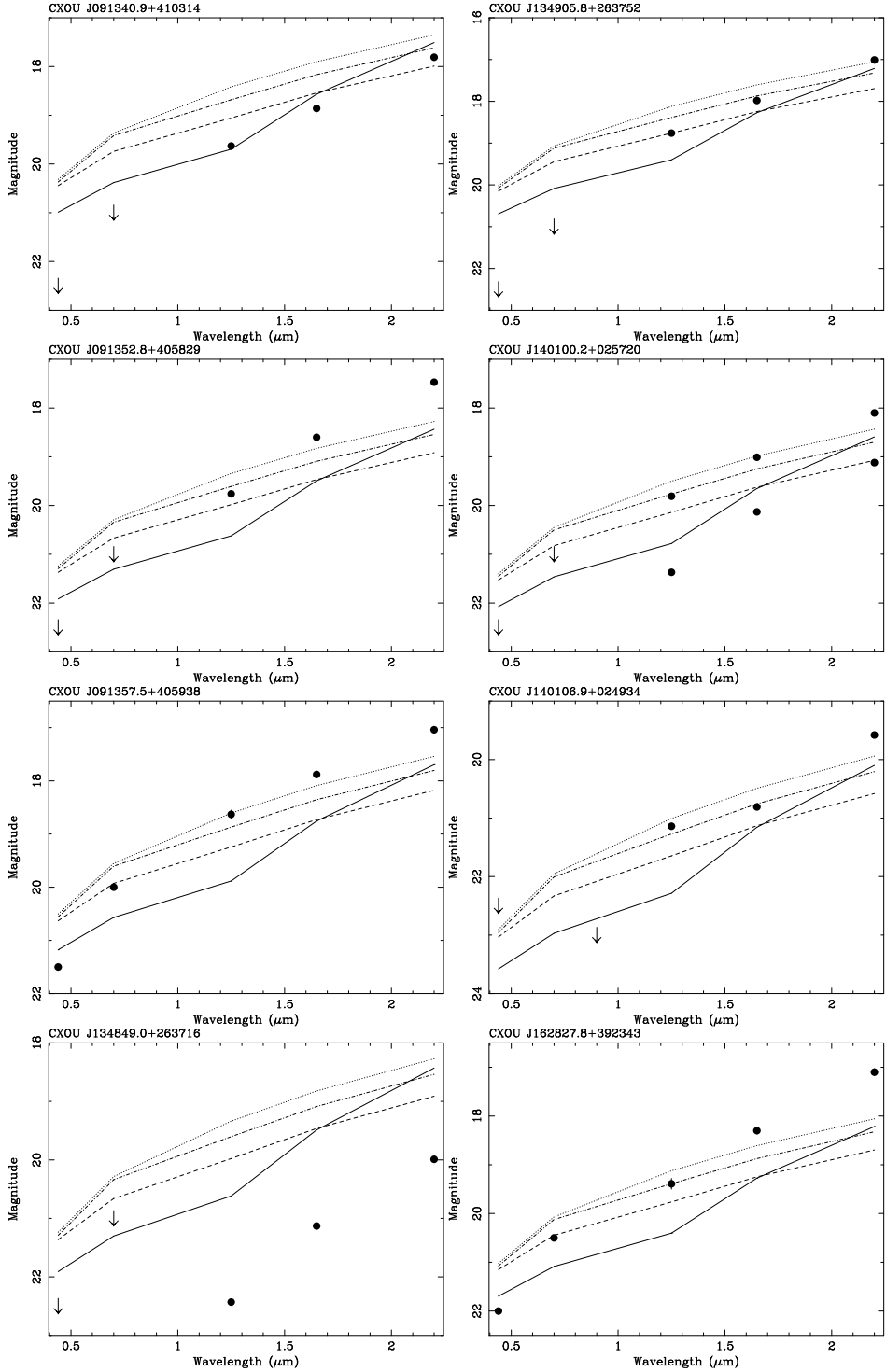

\vbox{
\hbox{  
\psfig{figure=fig7a.ps,width=0.35\textwidth,angle=270}
\psfig{figure=fig7e.ps,width=0.35\textwidth,angle=270}
}\hbox{
\psfig{figure=fig7b.ps,width=0.35\textwidth,angle=270}
\psfig{figure=fig7f.ps,width=0.35\textwidth,angle=270}
}\hbox{
\psfig{figure=fig7c.ps,width=0.35\textwidth,angle=270}
\psfig{figure=fig7g.ps,width=0.35\textwidth,angle=270}
}\hbox{
\psfig{figure=fig7d.ps,width=0.35\textwidth,angle=270}
\psfig{figure=fig7h.ps,width=0.35\textwidth,angle=270}
}}
\caption{\label{fig:seds}
Observed optical and infrared magnitudes (solid circle markers) for
the sources detected with TUFTI, plotted against those 
predicted from a typical PG quasar spectral energy
distribution. 
The lines mark magnitudes extrapolated into the optical-infrared
waveband from the observed X-ray flux for redshifts of 0$<z<$3 
($z=0$ solid, $z=1$ dashed, $z=2$ dash-dot and $z=3$ dotted
lines).
 }
\end{figure*}

\begin{figure}
\psfig{figure=fig8.ps,width=0.45\textwidth,angle=270}
\caption{\label{fig:sohrats}
Plots of $\Delta\chi^2$ = 2.3, 4.61 and 9.21 (corresponding to the
1-$\sigma$, 90 per cent and 99 per cent confidence regions for two
interesting parameters) as a function of redshift and intrinsic
absorption (N$_{\rm H}$) for the brightest X-ray source,
CXOU J134905.8+263752. }
\end{figure}

\begin{figure}
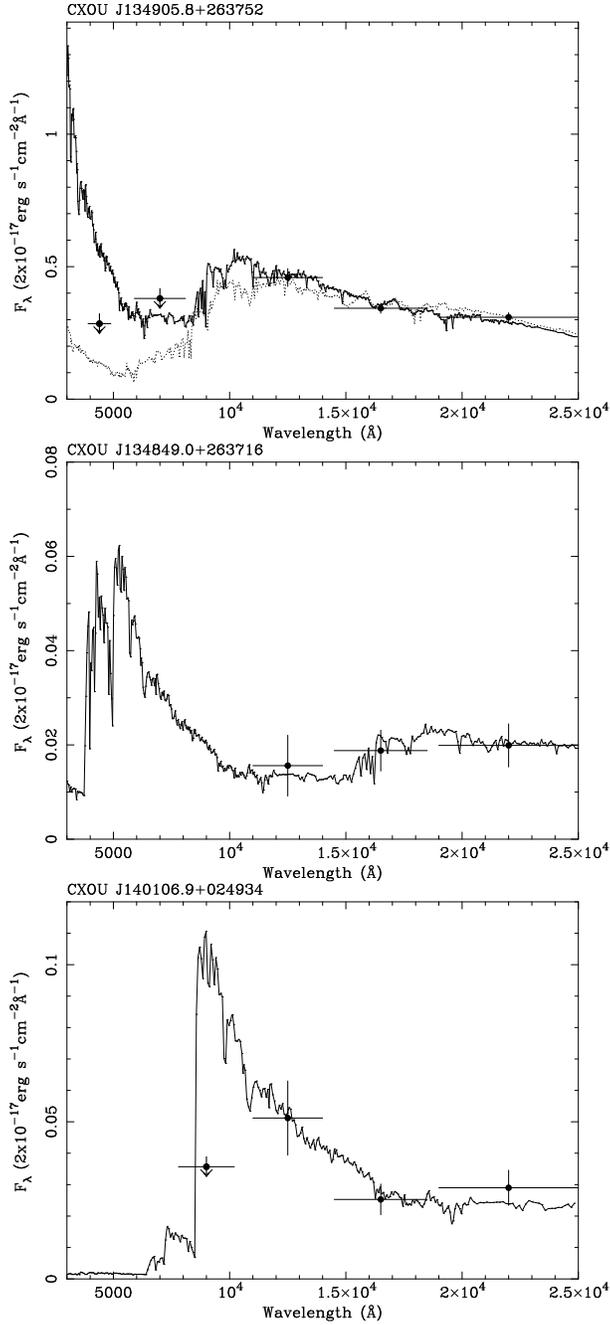

\vbox{
\psfig{figure=fig9a.ps,width=0.45\textwidth,angle=270}
\psfig{figure=fig9b.ps,width=0.45\textwidth,angle=270}
\psfig{figure=fig9c.ps,width=0.45\textwidth,angle=270}
}
\caption{\label{fig:photzfit}
Best-fit photometric solutions of a 3~Gyr-old elliptical galaxy
with no internal reddening (solid line) to the 
infrared magnitudes (solid circle markers) of CXOU
J134905.8+263752 (top), CXOU J134849.0+263716 (middle) and CXOU
J140106.9+024934 (bottom). 
The {\sl x}-errors on the infrared fluxes indicate the bandwidth. 
The solution for 
J134905.8+263752 is improved by either allowing a high intrinsic
absorption or an older
stellar population, such as the 5~G-yr old population with no internal
reddening shown by the dotted line in this plot). }
\end{figure}

\begin{figure}
\psfig{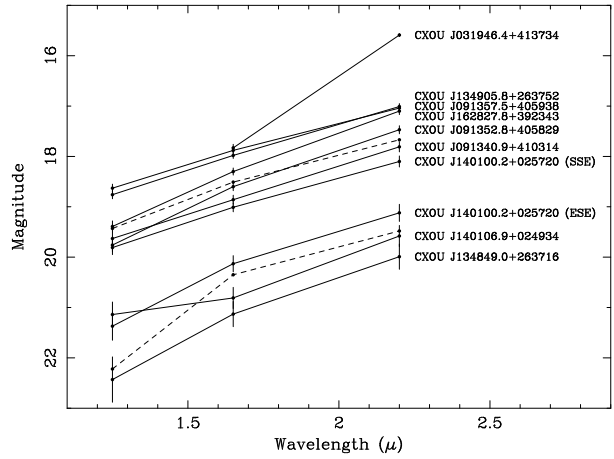}
\caption{\label{fig:magcomp}
J, H and K magnitudes of our detected sources (from
Table~\ref{tab:irmags}) compared to the expected magnitudes of a
passively evolving L$^*$ elliptical galaxy (R. McMahon \& A.
Aragon-Salamanca, private communication). The dashed lines indicate
the magnitudes of the L$^*$ galaxy at $z=2$ (lower) and $z=1$ (upper
line). }
\end{figure}

\onecolumn
\begin{figure*}
\vspace{-2.0cm}
%\hspace{-2.5cm}
\psfig{figure=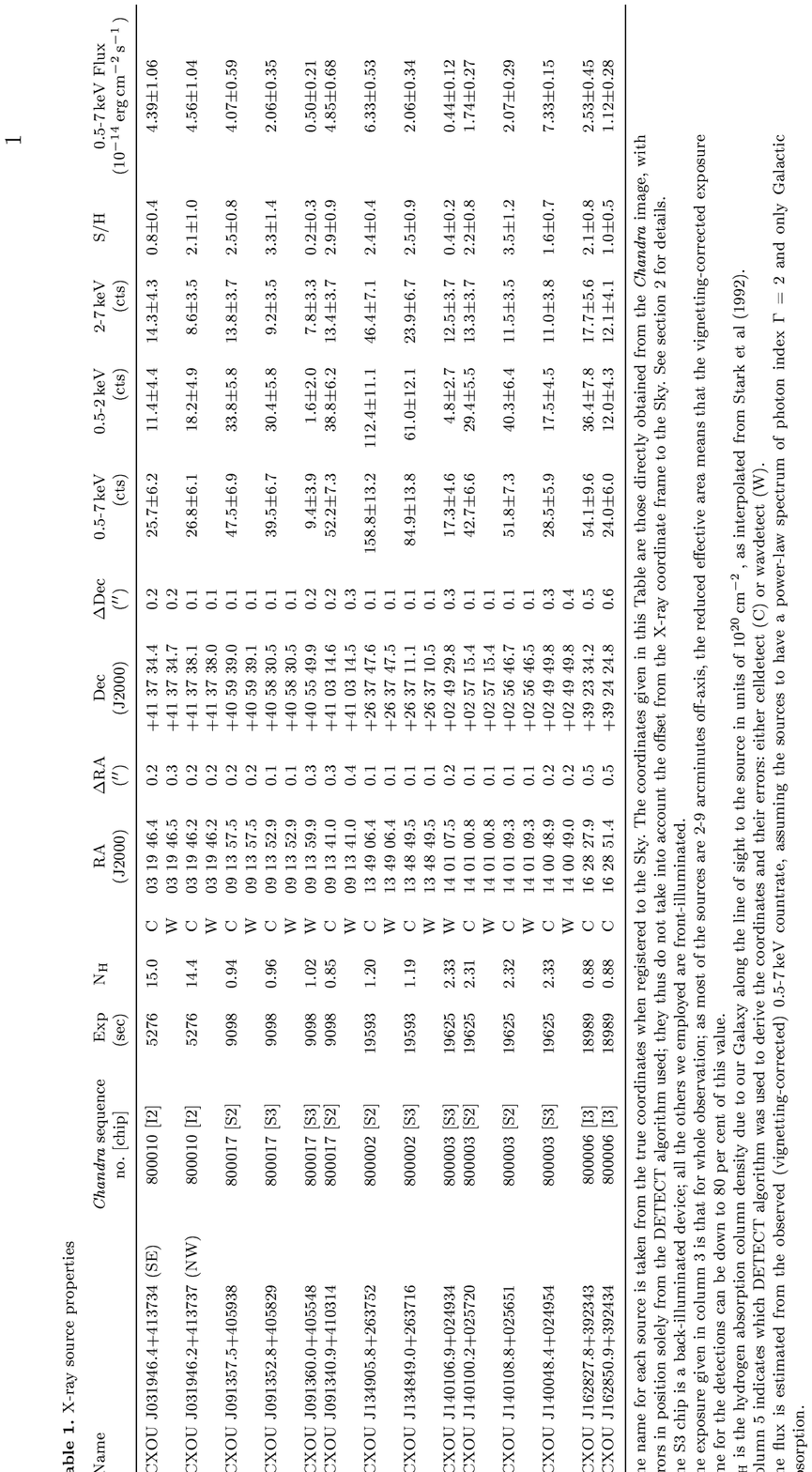,width=1.1\textwidth,angle=180}
\end{figure*}

\addtocounter{table}{+1}
\onecolumn
\begin{table}
\caption{Table of infrared observations \label{tab:irlog}}
\begin{tabular}{lccccccc}
           & & & & & & & \\
Name       & RA         & DEC      & \multicolumn{2} {c} {CGS4
Exposure (s) } &
\multicolumn{3} {c} {TUFTI Exposure (s) } \\
           & (J2000)    & (J2000)  & H    & K    & J      & H      & K       \\
% note that these are the directions TUFTI was pointed in, not the actual object info!
\hline 
CXOU~J031946.4+413734& 03 19 46.4 & +41 37 34& 1200 & 1920    & --- & --- & --- \\
%Perseus~wi1 & & & ---  & 960    & --- & --- & --- \\
% no longer include wi1 as no am sure that registration is not right
%-- mre detailed analysis suggests that xray core position is not
%centred on teh radio position and the registration may be incorrect
CXOU~J091340.9+410314 & 09 13 40.9 & +41 03 14 & ---  & ---  & 540 & 540 & 540 \\
CXOU~J091352.8+405829 & 09 13 52.7 & +40 58 30 & 1200    & 1200    & 540 & 540 & 540 \\
CXOU~J091357.5+405938 & 09 13 57.4 & +40 59 38 & ---  & 1080    & 540 & 540 & 540 \\
CXOU~J091360.0+405548 & 09 13 60.0 & +40 55 49 & ---  & 480    & 540 & 540 & 540 \\
CXOU~J134849.0+263716 & 13 48 48.9 & +26 37 14  & ---  & ---  & 540 & 540 & 540 \\
CXOU~J134905.8+263752 & 13 49 05.7 & +26 37 51 & 360    & 1080    & 540 & 540 & 540 \\
CXOU~J140048.4+024954& 14 00 48.4 & +02 49 53 & ---  & ---  & --- & --- & 540  \\
CXOU~J140100.2+025720& 14 01 00.1 & +02 57 19 & ---  & ---  & 360 & 360 & 360 \\
CXOU~J140106.9+024934& 14 01 06.8 & +02 49 33 & ---  & ---  & 540 & 540 & 540 \\
CXOU~J140108.8+025651& 14 01 08.8 & +02 56 50 & ---  & ---  & --- & --- & 540 \\
CXOU~J162827.8+392343& 16 28 27.8 & +39 23 42 & ---  & ---  & 540 & 540 & 540 \\
CXOU~J162850.9+392434& 16 28 50.9 & +39 24 34 & ---  & ---  & --- & --- & 540 \\
% need pg to work out final exposures for all the cgs4 data frames
% and to cross-check my exposure information for the ircam3 obs
% need to add seeing to this table? comment on air transparency? 
%                    & & & & & & & \\
\hline 
\end{tabular}
~~~~~~~~\par
The coordinates given are those for where TUFTI was aimed, except for
CXOU J031946.4+413734
where it refers to the CGS4 pointing position, which is the same as
the optical object associated with the X-ray source.  \\
Exposure times are given in seconds. \\
\end{table}

\begin{table}
\caption{Infrared magnitudes and offsets of objects from the 
X-ray position (ie the TUFTI pointing position). \label{tab:irmags}}
\begin{tabular}{lcrrrccc}
              && &  &           & & & \\
 Object       & RA offset&Dec offset& B & R &  J           & H        &    K \\%Mean Mags
              & (arcsec) & (arcsec) &  &           & & & \\
\hline
%CXOU J031946.4+413734  & --- & ---  & ???        & ???     & --- & 17.83$\pm$0.07$\dag$ & 15.59$\pm$0.03$\dag$ \\
CXOU J091357.5+405938 &0.77 &--0.19 & $\sim$21.5 &$\sim$20 & 18.63$\pm$0.08 &17.88$\pm$0.06 &17.04$\pm$0.06 \\
CXOU J091352.8+405829 &1.26 &--0.80 & $>$22.5&$>$21 & 19.76$\pm$0.13 &18.60$\pm$0.08 &17.47$\pm$0.08\\
CXOU J091340.9+410314 &0.69 &--0.54 & $>$22.5&$>$21 & 19.63$\pm$0.13 &18.86$\pm$0.09 &17.81$\pm$0.10\\
CXOU J134905.8+263752 &1.35 &  1.39 & $>$22.5 & $>$21 & 18.76$\pm$0.08 &17.98$\pm$0.06 &17.01$\pm$0.06\\
CXOU J134849.0+263716 &0.49 &2.04 & $>$22.5&$>$21 & 22.43$\pm$0.45 &21.13$\pm$0.25 &19.99$\pm$0.25\\
%            & 0.348& 5.69 & $>$22.5&$>$21  & 20.27$\pm$0.17 &19.60$\pm$0.12 &18.80$\pm$0.14\\
CXOU J140106.9+024934$\ddag$   &1.33 &1.20 & $>$22.5&$>$23$^*$  & 21.14$\pm$0.25 &20.81$\pm$0.21 &19.58$\pm$0.21\\
CXOU J140100.2+025720 (SSE)   & 2.23 &--2.48 & $>$22.5&$>$21  & 19.81$\pm$0.14 &19.01$\pm$0.09 &18.10$\pm$0.11\\
CXOU J140100.2+025720 (ESE)   & 4.48& --0.18 & $>$22.5&$>$21  & 21.37$\pm$0.28 &20.13$\pm$0.16 &19.12$\pm$0.17\\
CXOU J162827.8+392343 & 0.39 &0.73   &$\sim$22&$\sim$20.5 & 19.39$\pm$0.11 &18.30$\pm$0.07 &17.10$\pm$0.07\\
%              && &  &           & & & \\
\hline 
\end{tabular}
~~~ \par
The B and R magnitudes are estimated from the DSS, except for that
marked by $^*$, which is an I-band limit from a deep Palomar exposure of this field
(Smail, private communication). \par
%$\dag$ Magnitudes inferred from the H+K CGS4 spectrum. \par
$\ddag$ The infrared magnitudes are given for the main, brightest source of the
three resolved near the centre of the TUFTI image. \par
\end{table}

\begin{table}
\caption{Predicted ratios of soft (0.5-2\keV) to hard (2-7\keV) flux
for a power-law spectrum of given $z$ and N$_{\rm H}$. 
The differences between the predictions for the two sources reflect the different response matrices of the front-
(CXOU J134905.8+263752 on S2) and back- (CXOU J134849.0+263716 on S3)
illuminated CCDs. The observed values of the S/H are $2.4\pm0.4$ and
$2.5\pm0.9$  for CXOU J134905.8+263752 and CXOU J134849.0+263716
respectively. 
\label{tab:sohrats}}
\begin{tabular}{lclllll}
Source & N$_{\rm H}$ & \multicolumn{5} {c} {S/H ratios} \\
       & (\pcmsq)    &  $z=$0.1 & $z=$0.5 & $z=$1 & $z=$2 & $z=$3 \\
\hline 
CXOU J134905.8+263752 &$10^{21}$        &3.5	&3.8	&4.0	&4.2	&4.2\\
CXOU J134905.8+263752 &$10^{22}$	&1.2	&2.0	&2.8	&3.6	&3.9\\
CXOU J134905.8+263752 &$10^{23}$	&0.026	&0.17	&0.54	&1.5	&2.4\\
CXOU J134905.8+263752 &$10^{24}$	&1.9$\times10^{-3}$&3.7$\times10^{-3}$	&8.1$\times10^{-3}$	&0.10	&0.33\\
       & &  & & & & \\
CXOU J134849.0+263716 &$10^{21}$ &4.7	&5.3	&5.7	&6.0	&6.1\\
CXOU J134849.0+263716 &$10^{22}$ &1.3	&2.3	&3.4	&4.9	&5.5\\
CXOU J134849.0+263716 &$10^{23}$ &0.026	&0.18	&0.59	&1.7	&2.8\\
CXOU J134849.0+263716 &$10^{24}$ &6.2$\times10^{-5}$ &1.9$\times10^{-3}$ &6.7$\times10^{-3}$	&0.12	&0.36\\
\hline
\end{tabular}
~~~~~~\\
The  N$_{\rm H}$ given in column 2 is the column density {\sl
intrinsic} to the source. 
\end{table}

\begin{table}
\caption{Best-fit results from photometric redshift fitting for a
3~Gyr-old elliptical galaxy with no intrinsic reddening \label{tab:photzfit}}
\begin{tabular}{lcc}
 Object       & Redshift & Reduced $\chi2$  \\
              & $z$      &                  \\
\hline
CXOU J091357.5+405938  &1.185 (0.86,1.72) &0.600 \\
%    &b &1.185 (0.80,1.85) [0.18-0.54]&0.600 & 66.25 &0.0   &[3]\\
%    &c &1.185 (0.80,1.85) &0.600 & 66.25 &[0.0] &3\\
%    &d &1.185 (0.10,2.24) [4.62-6.0]&0.427 & 78.95 &0.60 &1.5\\
%    &e &1.810 &0.140 & 96.72 &1.80 &0.0132\\
CXOU J091352.8+405829 &3.100 (2.76,3.39) [2.02-2.51]& 2.020 \\
%    &b &1.325 (0.28,2.25) & 0.021 & 99.91 &1.20  &[3]\\
%    &c &1.865 (1.17,2.30) [2.93-3.31]& 1.021 & 39.48 &[0.0]  &5\\
%    &d &1.185 (0.28,3.70) [4.26-6.0]& 0.006 & 99.99 &0.90  &5 \\
%%    &e &5.220 (0.15,6.0)&0.000 & 100.00 &2.70& 0.0115\\
CXOU J091340.9+410314 &4.980 (4.68,5.48) [1.0-1.75]&0.496 \\
%    &b &6.000 (4.65,6.0) [0.76-1.91]&0.116 & 97.70 &0.60 &[3]\\
%    &c &1.220 (0.90,1.93) [4.24-5.66]&0.338 & 85.27 &[0.0] &5\\
%    &d &5.810 (3.16,6.0) [0.07-2.33]&0.018 &99.93 &1.20 &[1.5]\\
%%    &e &5.145 (&0.007 &99.99 &0.90 &1.0152\\
CXOU J134905.8+263752 &1.260 (1.00,1.65) [4.89-5.34]&1.713 \\
%     &b &6.000 (5.2,6.0) [0.85-1.66]&0.576 & 68.03 &0.60 &[3]\\
%    &c &1.100 (0.87,1.54)&0.568 & 68.59 &[0.0] &5\\
%    &d &5.370 (4.57,6.0) [0.13-2.0]&0.128 &97.23 &0.90 &1.5\\
%%    &e &5.070 & 0.027 &99.86 &1.20 &0.2550\\
CXOU J134849.0+263716 &3.085 (0.95,5.11)& 0.044 \\
%     &b &1.055 (0.00,6.00)& 0.001 & 100.00 &1.80 &[3]\\
%     &c &3.085 (0.35,6.00)& 0.044 & 99.64 &[0.0] &3\\
%     &d &0.995 (0.00,6.00)& 0.000 &100.00 &1.50 &5\\
%     &e &0.900 & 0.000 &100.00 &2.40 &1.434\\
CXOU J140106.9+024934 &5.995 (5.38,6.00) [0.94-1.92]&0.419 \\
%     &b &5.995 (4.78,6.00) [0.08-2.15]&0.265 &90.04 &0.30 &[3]\\
%     &c &5.995 (5.17,6.00) [0.23-2.22]&0.419 &79.52 &[0.0]& 3\\
%     &d &5.995 (0,6.0)&0.265 & 90.04 &0.30 &3\\
%     &e &5.915 &0.516 &72.37 &0.90 &1.0152\\
CXOU J140100.2+025720 SSE &1.695 (1.20,2.10)& 2.352\\
%     &b &6.000 (5.84,6.0) [0.97-1.98]& 0.961&42.76 &0.60 &[3]\\
%     &c &1.380 (0.99,1.93)& 1.080&36.45 &[0.0]& 5\\
%     &d &6.000 (5.46,6.00) [0.08-2.14]& 0.392&81.46 &1.20 &1.5\\
%     &e &6.000 & 0.184&94.67 &1.20 &0.5088\\
CXOU J140100.2+025720 ESE &2.225 (1.27,3.24)& 0.407\\
%     &b &1.800 (0.08,3.28) [5.21-6.0]& 0.053&99.48  &0.90 &[3]\\
%     &c &1.880 (0.82,3.29)& 0.081&98.81 &[0.0] &5\\
%     &d &1.695 (0.0-6.0)& 0.031&99.81 &0.30 &5 \\
%%     &e &6.000 & 0.024&99.88 &2.70 &0.0040\\
CXOU J162827.8+392343 &3.260 (2.98,3.44) [4.34-4.55]&1.811 \\
%     &b &3.260 (2.92,3.48) [0.86-2.28]&1.811  &12.35 &0.00 &[3]\\
%%     &b &3.260 (2.92,3.48) [0.86-2.28,4.27-4.62]&1.811  &12.35 &0.00 &[3]\\
%     &c &3.260 (2.93,3.69) [4.27-4.62]&1.811 &12.35 &[0.00]& 3\\
%     &d &3.410 (2.98,3.94) [0.89-2.29]&0.066 &99.20 &0.30 &1.5\\
%%     &e &3.420 &0.075 &98.97 &0.30 &1.4340\\
\hline 
\end{tabular}
~~~~\par
 In column 2, the 90 per cent (min-$\chi^2+2.7$) errors on the
redshift fit are given, and the range of redshift about any secondary
$\chi^2$ minimum is given within square brackets. \par
%In column 2, the full range for which $\chi^2\lt\chi^2+2.7$
%Summary of models: \par
%a) 3Gyr-old elliptical with no intrinsic reddening \par
%b) 3Gyr-old elliptical with intrinsic reddening\par
%c) a choice between an elliptical galaxy at ages of 1.5, 3 and 5Gyr,
%but with no intrinsic reddening \par
%d) as (c) but now with intrinsic reddening
%%e) a fully free fit of an elliptical galaxy of both varying age and reddening. \par
\end{table}

\end{document}